# SCIENTIFIC REPORTS

**OPEN**

# Generation of ultrahigh field by micro-bubble implosion

M. Murakami[1], A. Arefiev[2] & M. A. Zosa[1]

Breaking the 100-MeV barrier for proton acceleration will help elucidate fundamental physics and advance practical applications from inertial confinement fusion to tumour therapy. Herein we propose a novel concept of bubble implosions. A bubble implosion combines micro-bubbles and ultraintense laser pulses of $10^{20}$–$10^{22}$ W cm$^{-2}$ to generate ultrahigh fields and relativistic protons. The bubble wall protons undergo volumetric acceleration toward the centre due to the spherically symmetric Coulomb force and the innermost protons accumulate at the centre with a density comparable to the interior of a white dwarf. Then an unprecedentedly high electric field is formed, which produces an energetic proton flash. Three-dimensional particle simulations confirm the robustness of Coulomb-imploded bubbles, which behave as nano-pulsars with repeated implosions and explosions to emit protons. Current technologies should be sufficient to experimentally verify concept of bubble implosions.

Ion acceleration by intense lasers has been studied because the interaction between ultraintense ultrashort laser pulses and solid matter can produce energetic ions. Such ions have potential in numerous applications such as tumour therapy[1,2], radiography of dense targets[3], proton-driven inertial confinement fusion[4], and injection into conventional accelerators[5]. Additionally, the generation of high-energy protons is a goal in fields such as high-energy-density physics and astrophysics. The present paper provides a novel fundamental idea shedding light on an unexplored approach to generate unprecedentedly high fields and ion densities as well as the accelerated ion energy, that has never been proposed earlier.

Several schemes have produced energetic protons using high power lasers. Examples include target normal sheath acceleration[6–9], Coulomb explosion[10–16], radiation pressure acceleration[17–19], breakout afterburner acceleration[20,21], magnetic vortex acceleration[22,23], and collisionless shock acceleration[24–26]. Depending on the applied laser intensity, these schemes can generate protons with energies on the order of 10–100 MeV under applied laser intensities of $10^{20}$–$10^{22}$ W cm$^{-2}$. To date, only higher laser intensities have been considered to achieve higher proton energies.

Here we propose a new concept - bubble implosion. Suppose that spherical bubbles with radii of the order of $R_0 \simeq 0.1$–10 m are artificially contained in a uniform solid target, which is assumed in this paper to be pure hydrogen just for simplicity. When irradiating the target by ultraintense femtosecond laser pulse with an intensity of $I_L \simeq 10^{20}$–$10^{22}$ W cm$^{-2}$, hot electrons with temperature of $T_e \simeq 10$–100 MeV are generated according to the Ponderomotive scaling[6]. The hot electrons run around in the target to ionise the atoms to the ionization state $Z = 1$ almost instantaneously with its initial solid density $n_{i0} \simeq 5 \times 10^{22}$ cm$^{-3}$ being kept constant. The hot electrons fill the bubbles in a very short period, the characteristic time of which is $R_0/c \lesssim$ a few fsec (Fig. 1(a)), where $c$ is the speed of light. It should be noted that the high mobility of hot electrons often result in unwelcome energy dissipation and entropy increase in many applications. However, in the present scheme, such features of electrons play the crucial role to provide super high uniformity of the implosion and an ultrahigh field.

Because of the electrons flying in the bubble, the ions on the bubble surface "feel" strong electrostatic (Coulomb) force and begin volumetric implosion toward the bubble centre as illustrated in Fig. 1(b). The innermost ions continue to implode until they are unprecedentedly compressed to a nanometer scale such that their radial inward motion is stopped by the resulting outward electric field. Upon collapse of the bubble, the innermost ions "find" that their following ions just behind them have built up an extraordinary steep slope of Coulomb potential. Then they slide down the slope with resulting energies many times higher than the energy gained during the implosion. Figure 1(c) illustrates the envisioned mechanism with the main events depicted on the same image, i.e., laser illumination, hot electron spread, bubble implosion, and proton flash. One might expect that such an "anomalous" ion acceleration may occur in laser interaction with porous materials like foam[27]. However, as apparently understood from a simple model given below, the special characteristics of bubble implosion can be realized under high symmetry of the hollow and surrounding nanostructures.

[1]Institute of Laser Engineering, Osaka University, Osaka, 565-0871, Japan. [2]UC San Diego, 9500 Gilman Drive, La Jolla, CA, 92093-0411, USA. Correspondence and requests for materials should be addressed to M.M. (email: murakami-m@ile.osaka-u.ac.jp)

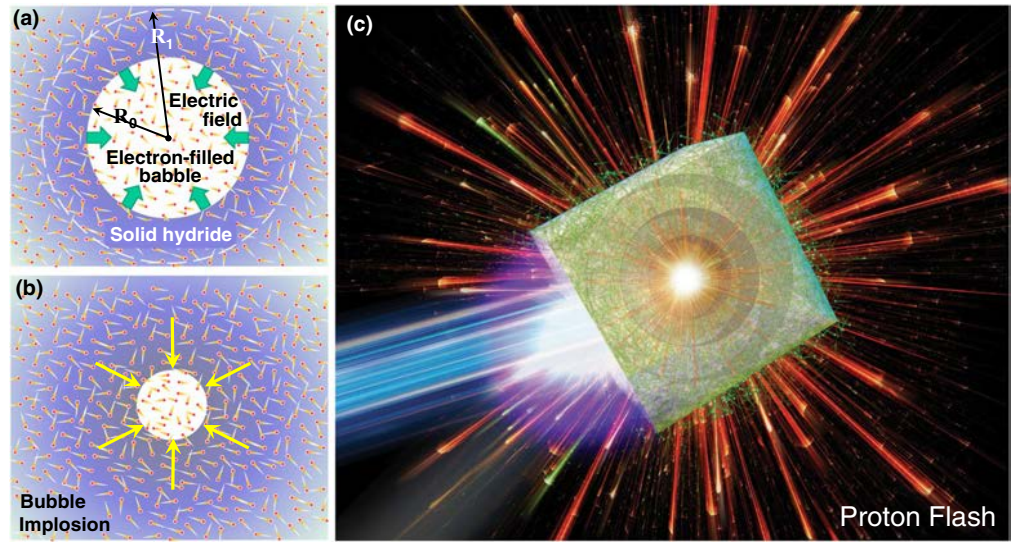

**Figure 1.** (**a**) Initial phase: The inner volume of a bubble surrounded by solid matter is filled with hot electrons. Bulk atoms composing the solid matter will be ionised in a very short time to "feel" the volumetric Coulomb force because of the bubble electrons. (**b**) bubble implosion phase: The ions begin to accelerate toward the bubble centre in a spherically symmetric manner, where the innermost ions are most strongly accelerated until the very final moment of the collapse at the centre. (**c**) Envisioned picture showing the whole main events integrated, i.e., laser illumination, hot electron spread, implosion, and proton flash.

Phenomena such as converging shock waves[28] and sonoluminescence[29] are similar to a bubble implosion. Shock waves are observed in many branches of physics. Although sonoluminescence is a relatively new phenomenon in the acoustics field, Lord Rayleigh proposed the basic idea (contraction of a water bubble) over a century ago[30]. The behaviour of bubble implosions reported in this study remarkably differs. Extremely high temperatures and low densities characterise the physical states of collapsing converging waves at the centre in shock waves and sonoluminescence. By contrast, extremely high densities and practically zero temperatures for protons characterise bubble implosions.

It should be also noted that Nakamura *et al.*[31,32] have reported another seemingly similar phenomenon - "Coulomb implosion". However, the two implosions are phenomenologically different from each other. The Coulomb implosion occurs for negative ions with a much smaller fraction than the bulk positive ions, that are expanding via Coulomb explosion after the most electrons are blown off by an intense laser. On the other hand, the bubble implosion is driven by the bubble-filled electrons. Because of the essential difference in the electrons role, the bubble implosion results in, for example, an overwhelmingly higher compressed density of ions than Coulomb implosion.

We postulate below that the total volume of bubbles, $V_b$, is much smaller than the volume of a solid matter containing the bubbles, $V_s$. In other words, the bubble disoccupancy $\alpha$ is close to unity, i.e., $\alpha \equiv 1 - V_b/V_s = 1 - (R_0/R_1)^3 \simeq 1$, where $R_1$ stands for a virtual radius of a spherical solid assigned to a single bubble (Fig. 1(a), dashed circle). Also we postulate that the electron temperature $T_e$ is so high that the electrons distribute uniformly over the whole region of the target. This isotropic assumption of electron distribution will be demonstrated in the PIC simulation given later. The total electric charge integrated from the centre to an arbitrary radius $r$, i.e., $Q(r) = \int_0^r 4\pi r^2 e(Zn_i - n_e)dr$ with $e$ being the electron charge, and the local electric field, $E_f(r)$, are related by Gauss's law in the form, $E_f(r) = Q(r)/r^2$. The whole ions thus begin to accelerate inward according to the electric field. The maximum implosion velocity is achieved when they reach at around the centre, which corresponds to the energy done by the electric field, $\mathcal{E}_0 = \int_0^{R_0} eE_f(r)dr$, given by

$$\mathcal{E}_0 = \frac{e^2 N_{e0}}{2R_0} \simeq 15\,\text{MeV} \cdot \left(\frac{\bar{n}_{e0}}{5 \times 10^{21}\,\text{cm}^{-3}}\right)\left(\frac{R_0}{1\,\mu\text{m}}\right)^2, \tag{1}$$

where $N_{e0} = (4\pi/3)R_0^3 \bar{n}_{e0}$ is the total electron number contained in the initial bubble with $\bar{n}_{e0}$ being the average electron density in the bubble. Note that $\bar{n}_{e0} \approx n_{i0}/10 \approx 5 \times 10^{21}\,\text{cm}^{-3}$ is here employed as a reference value, which is actually in the same order as those obtained in numerical simulations discussed later. It is shown below that $N_{e0}$ is the one and only crucial "extensive" variable, essentially differentiating the present scheme from the aforementioned schemes.

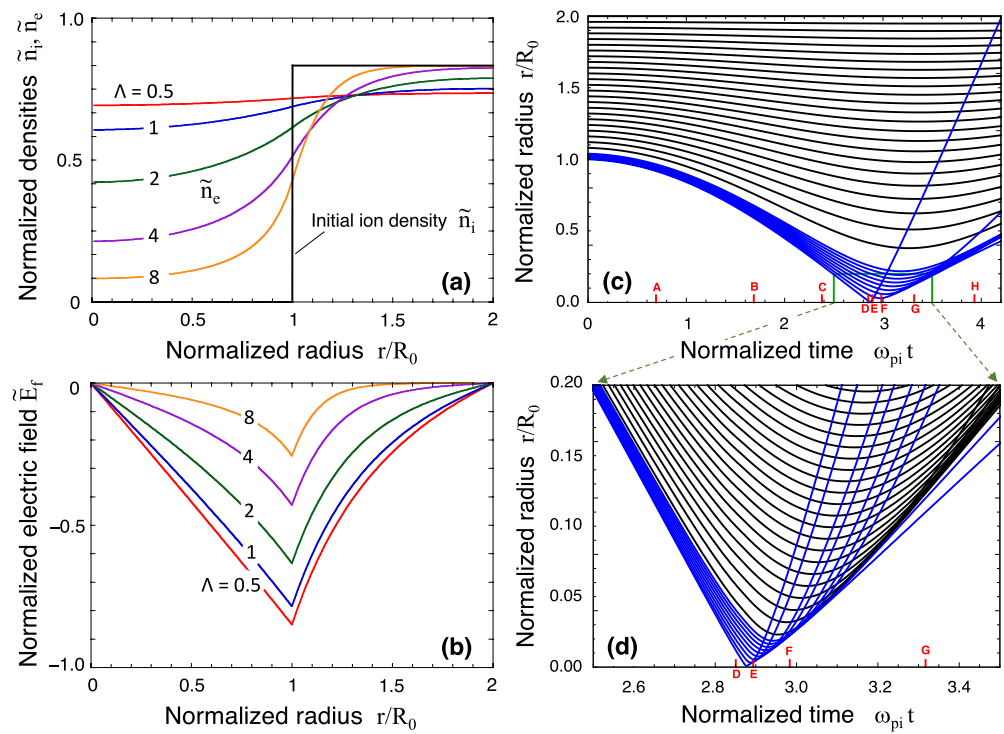

**Figure 2.** (**a**) Initial density profiles of the electron $\tilde{n}_e = n_e/n_{i0}$ and the ion $\tilde{n}_i = n_i/n_{i0}$ normalised by the initial ion density $n_{i0}$, respectively, and (**b**) the electric field $\tilde{E}_f = E_f[(4\pi/3)R_0 n_{i0} Ze]^{-1}$ as a function of $\Lambda \equiv R_0/\lambda_{De}$ under $R_1/R_0 = 2$ and $Z = 1$. (**c**) Ion trajectories obtained by 1D simulation under $R_1/R_0 = 2$ and $\Lambda = 0.5$. The black curves correspond to initial radii with constant increment of $\Delta r/R_0 = 0.04$, while the blue curves divide the innermost segment corresponding to $r/R_0 = 1.0 \pm 1.04$, to better observe the implosion dynamics. The labels, A - H, are to compare other physical quantities in the following Figures. For the solid density, the normalised unit time corresponds to 3.4 fsec. (**d**) Magnified view of the rectangle part in (**c**). The total number of the flashed protons is nearly equal to 0.3±0.5% that of the electrons contained in the initial bubble volume.

To extract the salient features, we conducted 1D simulations of the bubble implosion. For simplicity, the electrons are assumed to obey the Boltzmann relation, because the electron mass is significantly smaller than an ion. The electrons are then described by Poisson-Boltzmann (P-B) equation, $\nabla \cdot \nabla \phi = 4\pi e[n_{ec} \exp(e\phi/T_e) - n_i]$, where $\phi$ is the electric potential and $n_{ec}$ is the temporal electron density at the centre. The P-B equation is furthermore simplified as a function of the dimensionless parameter $\Lambda \equiv R_0/\lambda_{De}$, where $\lambda_{De} = \sqrt{T_e/4\pi \bar{n}_{e0} e^2}$ is the Debye length. The parameter $\Lambda$ characterises to which extent the bubble is filled with the electrons. As a function of $\Lambda$, the P-B equation is numerically solved to give $\phi(r)$ and thus $n_e(r)$ under the appropriate boundary conditions. The electric potential profile is thus determined at every time step according to the ion motion in the field.

The ions are computed by particle-in-cell (PIC) method. Furthermore, only in the very limited central volume for $r \lesssim 0.02 R_0$, the ion motion is calculated based on the scheme of molecular dynamic (MD) simulation. This is because the tiny central region is the key domain where an ultrahigh field is formed to generate high energy protons. Therefore one needs to precisely evaluate the protons dynamics instead of using the averaged field prescribed by the PIC method. In the present 1D simulation in spherical geometry, we employed 2000 fixed grids, and $2 \times 10^4$ pseudoparticles.

Figure 2(a,b) shows the initial profiles for the electron density $n_e(r)$ and the electric field, respectively, obtained for different values of $\Lambda$ and a fixed initial ion density profile normalised by $n_{i0}$ under $R_1/R_0 = 2$. The electron profiles for $\Lambda \lesssim 1$ are rather flat over the entire domain, while they conspicuously reduce in the bubble with increasing $\Lambda (\gtrsim 2)$. It is convenient to normalize time $t$ and use the dimensionless quantity $\omega_{pi0} t$ instead, where $\omega_{pi0} \equiv \sqrt{4\pi n_{i0} Z^2 e^2/m_i}$ is the ion plasma frequency and $m_i$ is the ion mass. As a reference, $\omega_{pi0} \simeq 3.4$ fsec$^{-1}$ for solid density protons with $n_{i0} = 5 \times 10^{22}$ cm$^{-3}$.

Figure 2(c) shows the ion trajectories for the entire time region under the bubble conditions of $R_1/R_0 = 2$ and $\Lambda = 0.5$. The black curves correspond to initial radii with a constant increment of $\Delta r = 0.04 R_0$, while the blue curves subdivide the innermost segment to better resolve the implosion dynamics. The labels along the time axis, A - H, are to compare other physical quantities in subsequent figures. Figure 2(d) shows a zoom-in of the rectangular in Fig. 2(c). Until time D, all of the ion trajectories remain laminar, so that one curve does not intersect another. However, upon the collapse (time E), the innermost trajectory is strongly ejected radially outwards and this is the phenomenon that we call the proton flash. In Fig. 2(d), the innermost seven trajectories in blue represent flashed protons and they behave quite differently from the other trajectories. These trajectories sharply cut across the other trajectories, confirming that the flashed protons quickly slide down a Coulomb potential that can be effectively viewed as quasi-static. These "runaway" protons are emitted from a very small volume with $r \lesssim 0.05 R_0$ due to an explosive acceleration under the ultrahigh electric field that is generated by the accumulated proton core at the centre.

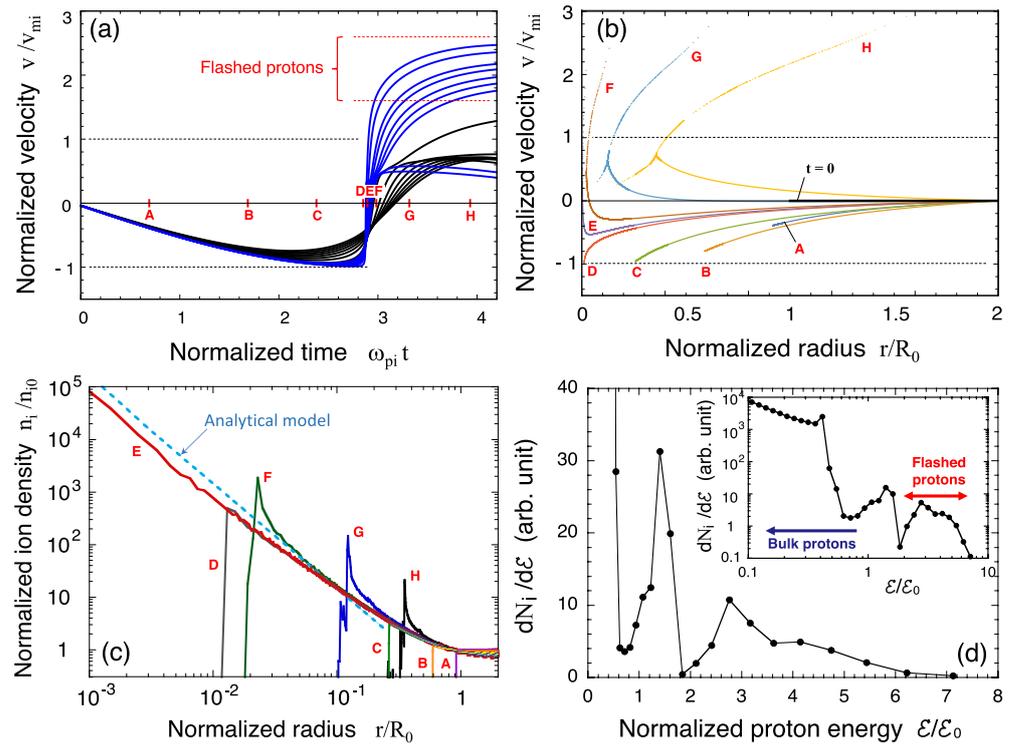

**Figure 3.** (**a**) Temporal evolution of the velocities of individual protons normalised by the maximum implosion velocity $v_{mi}$. The set of blue curves correspond to that in Fig. 2(d), while the black curves are appropriately extracted out of those in Fig. 2(d) for better observation. (**b**) Overall view of the velocity profiles at the snapshot times A - H. (**c**) Density profiles normalised by the solid density $n_{i0}$ at different times A - H. (**d**) Energy spectrum at time H. The inset stands for the same data given in the main frame but in double-logarithmic scales. The unit of the energy is chosen to be the maximum kinetic energy in the implosion phase, $\mathcal{E}_0$. The two-humped structure for $\mathcal{E}/\mathcal{E}_0 > 1$ is attributed to the acceleration process at the singular behaviour at the centre.

Figure 3(a) shows the velocity evolution of the flashed protons and the surrounding protons, normalised by the maximum implosion velocity $v_{mi}$. The blue and black curves correspond to those in Fig. 2(d). Upon the collapse (times D - F), the velocity of the flashed protons drastically increases, exceeding the maximum implosion velocity by a factor of 2.0±2.5, which are simply squared to give corresponding energy amplification by a factor of 4÷6. This energy amplification for the flashed protons is due to their sliding down the steep Coulomb potential slope. The innermost protons are the first ones to be reflected near the centre. The dynamics of the other protons that follow and that are located a bit further outwards is similar, but the expulsion is slightly delayed and the resulting energy amplification factor is smaller.

Figure 3(b) shows snapshots of the proton velocity as a function of radius for times A - H in Fig. 2(c). For practical laser and target parameters, the proton flash occurs over a very short time interval ($\lesssim$0.5 fsec) and a very small volume ($\lesssim$ a few nm) corresponding to times D - F. The flashed protons have a much higher velocity than surrounding bulk protons, as can be seen in snapshots G and H. It should be noted that, at such later times, a snowplow-like two-stream-structure is formed. The difference in velocity between the two streams is of the order of $v_{mi}$. In this simulation, the total number of flashed protons is found to be roughly $10^{-2} N_{e0}$.

We developed a simple model to understand what determines the fraction of the flashed protons and the corresponding energy amplification factor, assuming $\alpha \simeq 1$ and $\Lambda \lesssim 1$. In other words, the electrons maintain an almost uniform density in the entire system at $\bar{n}_e \approx n_{i0}$. Motivated by the already presented simulation results, we assume that the protons keep their order in space without overtaking each other (or nonbreaking) until the moment of the collapse. Note that this nonbreaking assumption of ion flow was also employed in refs[31,32]. Under these assumptions, the equation of motion for a proton prior to the collapse, $m_i \ddot{r} = eQ(r)/r^2$, that was located at $r(t=0) = r_0$, is given by

$$\frac{1}{\omega_{pi0}^2}\frac{d^2r}{dt^2} = \frac{r_1^3}{3r^2} - \frac{r}{3}, \quad (2)$$

where $r_1 \equiv (r_0^3 - R_0^3)^{1/3}$. Physically the first and the second terms on the right-hand side of Eq. (2) indicate that the Coulomb force is due to the protons and the electrons contained in the volume at radii smaller than $r(t)$, respectively. The maximum implosion speed of a proton at an initial position $r_0$ occurs at $r = r_1$ when $\ddot{r} = 0$. Analytical integration of Eq. (2) provides the velocity of a proton, $v = \dot{r}$, as a function of position $r$ as

$$\frac{v^2}{\omega_{pi0}^2} = \frac{2}{3}\left(\frac{1}{r_0} - \frac{1}{r}\right)r_1^3 + \frac{1}{3}(r_0^2 - r^2). \quad (3)$$

Equation (3) indicates that the maximum implosion speed $v_{mi} = \omega_{pi0} R_0 / \sqrt{3}$ is achieved by the innermost protons $r_0 = R_0$ as they reach the centre. Additionally, a proton with its initial position $r_0$ is halted ($v = 0$) due to the Coulomb repulsion at $r_2 \equiv \left(\sqrt{9 - 8(R_0/r_0)^3} - 1\right)r_0/2$. Using $dr_2/dr_0 \simeq 6$ for $r_0 \approx R_0$ derived from above analysis, the mass conservation $n_2 r_2^2 dr_2 = n_{i0} r_0^2 dr_0$ is reduced to give the density profile of the innermost protons upon the collapse as

$$n_i(r) = \frac{n_{i0}}{6}\left(\frac{R_0}{r}\right)^2, \quad (4)$$

where the subscript "2" is dropped for simplicity.

Figure 3(c) shows how the proton density evolves in time in the presented 1D simulation. Upon the collapse (time E), the density in the innermost grid exceeds the original density by five orders of magnitude, with $n_i/n_{i0} > 10^5$. This extraordinarily high compression shown in Fig. 3(c) has a power-law dependence, $n_i \propto r^{-2}$, which agrees with Eq. (4) depicted as the dashed line. It is worth pointing out that, in an agreement with our assumptions, the electron density indeed remains almost flat throughout the whole process due to the high electron mobility. After the proton flash, bulk ions also rebound to expand outward. A salient feature of this stage is the peaked structure in the density profile that is formed on the expanding bubble surface (times F - H).

The ion energy spectrum $dN_i/d\mathcal{E}$ at time H is shown on double-linear scales (Fig. 3(d)) and double-logarithmic scales (inset). The energy is normalised to the maximum kinetic energy, $\mathcal{E}_0$, defined by Eq. (1). The energy amplification for the flashed protons, $\mathcal{E}/\mathcal{E}_0$, ranges from 3 to 7. The two-humped structure for $\mathcal{E}/\mathcal{E}_0 \gtrsim 1$ is attributed to a complex behaviour of the innermost ions as their trajectories overlap upon the collapse.

The minimum radius $r_{min}$ achieved at the maximum compression is determined by the dynamics of the innermost ions. The compression stops when the maximum kinetic energy, $\mathcal{E}_{kin}$, gained during the implosion is converted into the potential energy, $\mathcal{E}_{pot}$. The total kinetic energy of the innermost ions is $\mathcal{E}_{kin} = \frac{1}{2} N_a m_i v_{mi}^2$, where $N_a$ is the total number of these ions. Considering that the characteristic interatomic distance on the initial bubble surface is $d_0 = n_{i0}^{-1/3}$, we find $N_a = 4\pi R_0^2/d_0^2$. Meanwhile, the potential energy is $\mathcal{E}_{pot} = \frac{1}{2}(N_a e)^2 r_{min}^{-1}$. Using the energy balance, $\mathcal{E}_{kin} = \mathcal{E}_{pot}$, we find $r_{min} = N_a e^2/m_i v_{mi}^2 = 3 n_{i0}^{2/3}/n_{e0}$. For example, this gives $r_{min} \simeq 0.81$ nm for $n_{i0} = 5 \times 10^{22}$ cm$^{-3}$. It should be noted that $r_{min}$ depends only on the initial ion density $n_{i0}$, and not the initial radius $R_0$ or the ionization state $Z$. At $r = r_{min}$, the maximum ion density $n_{max}$ is given with the help of Eq. (4) as

$$\frac{n_{max}}{n_{i0}} = \left(\frac{N_{e0}}{6^{7/2}\pi}\right)^{2/3}. \quad (5)$$

The maximum radius where the scaling (4) is applicable can be roughly estimated by solving, $n_i(r) = \bar{n}_i \approx n_{i0}$, to give $r_{max} = R_0/\sqrt{6}$. The applicable range for the derived density scaling is then approximately defined by $r_{min} \lesssim r \lesssim r_{max}$. The numerical factors in the expressions for $r_{min}$ and $r_{max}$ are not significant, because the energy amplification factor that is derived below is only logarithmically sensitive to $r_{min}$ and $r_{max}$. For example, for $R_0 = 2$ m and $\bar{n}_{e0} = n_{i0}/10 \approx 5 \times 10^{21}$ cm$^{-3}$, Eq. (5) gives $n_{max}/n_{i0} \approx 2 \times 10^5$, which is comparable to interior densities of a white dwarf. Note that, under such ultrahigh densities of hydrogen isotopes, pycnonuclear reaction can be discussed as a potential application[33].

Using the compressed density profile given by Eq. (4), we can now find the corresponding profile of the radial electric field $E_f$ that causes the proton flash. The centrally condensed positive charge, $Q(r) = \int_0^r 4\pi r^2 e n_i(r) dr = Q_0 r/2R_0$, readily yields $E_f(r) = Q(r)/r^2 = Q_0/2R_0 r$, where $Q_0 = N_{e0} e$ is the total electron charge contained in the initial bubble. In a Coulomb imploded core, the electric field is higher as the radius decreases. Here the electron contribution is neglected when evaluating $Q(r)$ near the centre, because $n_i \gg n_e$ in the highly compressed ion core. These radial dependencies remarkably differ from those for the well-known classical case of a uniformly charged sphere (or Coulomb explosion) with $n_i(r) = const$, i.e., $Q \propto r^3$ and $E_f \propto r$. For example, assuming $\bar{n}_{e0} = 5 \times 10^{21}$ cm$^{-3}$, $R_0 = 2$ m, and $r = 1$ nm as a characteristic scale of the core, the above scaling for a bubble-imploded core predicts $E_f \approx 6 \times 10^{14}$ V/m. This value of the electric field is roughly three orders of magnitude higher than the fields observed in current laser-plasma experiments, and six orders of magnitude higher than the maximum accelerating field achieved in the conventional accelerators driven by radio frequency (RF) fields[6].

Next we evaluated the energy amplification factor. The maximum kinetic energy $\mathcal{E}_{max}$ of flashed protons corresponds to the Coulomb potential gap that has been built up around the bubble centre when the innermost ions start to expand, i.e., $\mathcal{E}_{max} \approx \int_{r_{min}}^{r_{max}} eE_f(r)dr$, which is reduced to give the energy amplification factor as

$$\frac{\mathcal{E}_{max}}{\mathcal{E}_0} = \frac{1}{3}\ln\left(\frac{N_{e0}}{6^{7/2}\pi}\right). \quad (6)$$

For example, Eq. (6) gives $\mathcal{E}_{max}/\mathcal{E}_0 \simeq 4.2 \pm 6.5$ under $R_0 \simeq 0.3 \pm 3$ m and $\bar{n}_{e0} = 5 \times 10^{21}$ cm$^{-3}$, which agrees well with the 1D simulation results presented in Fig. 3(a,b).

We also estimated the number of highly accelerated protons as flashed protons. For simplicity, flashed protons are considered to be protons exceeding the threshold, which is defined as half maximum energy, i.e., $\mathcal{E}_{1/2} \equiv \frac{1}{2}\mathcal{E}_{max}$. Using Eqs (4±6), the normalised total number of protons with energies $\mathcal{E} \geq \frac{1}{2}\mathcal{E}_{max}$ is given by

$$\frac{N_{1/2}}{N_{e0}} = \left(\frac{6^{1/2}\pi}{64 N_{e0}}\right)^{1/6}. \quad (7)$$

For example, $N_{1/2}/N_{e0} = 0.9\%$ for $R_0 = 2$ m and $\bar{n}_{e0} = 5 \times 10^{21}$ cm$^{-3}$, which is close to the 1D simulation result in Fig. 3(d).

To investigate the bubble implosion in more detail, we conducted 3D simulations. This is a distinctly multi-scale problem, since both the spatial and temporal scales of a bubble implosion vary over four orders of magnitude from 1 nm to 10 m and from 0.01 fsec to 100 fsec, respectively. We used both particle-in-cell (PIC) and molecular dynamics (MD) approaches to tackle this challenging problem. PIC simulations can provide a comprehensive physical picture by treating a lot of particles, but the dynamic range is limited because of the fixed size of the cartesian cells. In contrast, MD simulations can treat the dynamics over a much wider dynamic range, taking all binary collisions into account, but because of that they are limited to a much smaller number of particles. In what follows, we complementarily use PIC and MD simulations to examine global features of the phenomenon and the localized behaviour of the innermost protons during the bubble implosion, respectively.

3D $(x, y, z)$ PIC simulations were conducted with open-source fully relativistic code EPOCH[34] using the periodic boundary conditions for particles and fields, while placing the bubble into the middle of the cubic computational domain. This approach simulates having multiple equally spaced bubbles inside the considered heated material. We set the cell sizes at 2 nm, because the key physical events of a bubble implosion occur on a nanometre-scale. The box size must be more than double the diameter of the bubble to ensure a spherically symmetric implosion and to avoid interference from neighbouring bubbles. The computational domain size was $240 \times 240 \times 240$ nm$^3$, while the initial bubble radius was $R_0 = 60$ nm.

Figure 4(a) shows snapshots of the bubble implosion at different times obtained by the 3D-PIC simulation, where the density distributions on the $x$-$y$ plane are colour-coded. At $t = 0$, the bubble in the middle of the box is empty. We initialised an otherwise uniform proton plasma composed of hot electrons and cold ions with $T_e = 1$ MeV, $Z = 1$, and $n_{i0} = n_{e0} = 3 \times 10^{21}$ cm$^{-3}$, from which the period for one cycle is estimated to be $T_{cyc} = 2\pi/\omega_{pi0} = 87$ fsec. The hot electrons quickly fill in the bubble volume and the implosion is launched. After the bubble collapses at $t \approx 45$ fsec $\approx (1/2)T_{cyc}$ (first flash), the bubble expands and then shrinks again to show a second flash at $t \approx 130$ fsec $\approx (3/2)T_{cyc}$ and then a third flash. This oscillating behaviour is confirmed by the 1D simulations. The bubble thus behaves as a nano-pulsar, alternating implosions and explosions to periodically emit energetic protons. The highly robust bubble oscillation is attributed to the collective nature of the spherically symmetric Coulomb system. Here it should be noted that the converging flows are generally unstable. As a matter of fact, the azimuthally asymmetric modes growing in time are well seen in Fig. 4(a). This should impose a constraint on the achievable ion energy. Also it should be noted that the square-shaped compressed core in panel 10 is likely to be attributed not only to the physical reason but also to the Cartesian mesh scheme employed in the PIC code. It is however beyond our scope in this paper to discuss in detail how the degraded sphericity of the bubble or the mesh structure affect the bubble implosion performance.

Figure 4(b) shows the time evolution of the proton energy spectrum. The maximum compression of the protons is $n_{max}/n_{i0} \simeq 350$, which is 2.5 times larger than $n_{max}/n_{i0} \simeq 140$ predicted by Eq. (5). This difference may be because proton convergence to the centre in three-dimensions may have discrepancies in time and space compared to the perfect 1D model. On the first flash, the maximum values of the implosion energy and the reflected proton energy are respectively read off to be $\mathcal{E}_0 \simeq 25$ keV and $\mathcal{E}_{max} \simeq 150$ keV, and thus the energy amplification factor $\mathcal{E}_{max}/\mathcal{E}_0 \simeq 6$, which agrees well with the 1D prediction. After the first flash, the maximum proton energy further increases on the successive flashes, though only weakly. Overall, it turns out that the 3D-PIC simulation results and the simple model agree qualitatively.

The isotropic behavior of the electrons assumed in the simple model and the 3D simulation plays a crucial role in the concept of bubble implosion. Meanwhile, illumination of an ultrashort ultraintense laser on matter generally produces extremely violent electromagnetic fields and resultant complex plasma motion. To demonstrate that a symmetric bubble implosion can indeed be achieved under an asymmetric laser-matter interaction circumstance, therefore, we conducted another simulation, which is still primitive but acceptably realistic. Since one needs then wider interaction space than earlier, we conducted 2D $(x, y)$ PIC simulation with the computational domain size of $1200 \times 1200$ nm$^2$ and the cell size of 1 nm, and thus keeping the total computational size in the same order as in the 3D case. Here it should be noted that cylindrically symmetric bubble (column) implosions can also be discussed in a similar manner to the spherical case, though we do not discuss it in this paper.

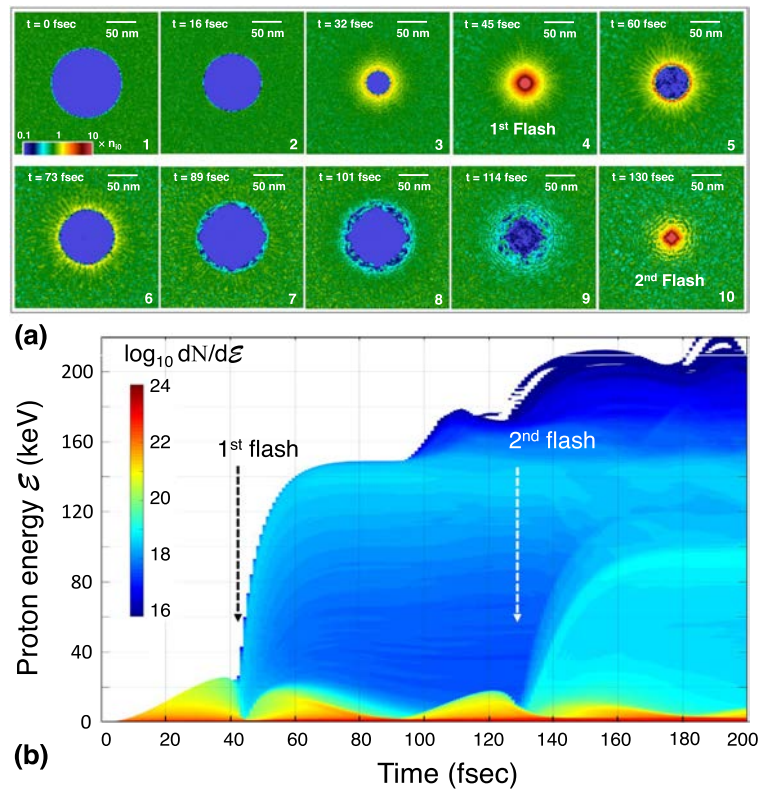

**Figure 4.** Three-dimensional ($x$, $y$, $z$) PIC simulation result using EPOCH code for a proton plasma. Laser-matter interaction is not taken into account here for simplicity. Each side of a periodic cube box and the cell size are set to be 240 nm and 2 nm, respectively. At $t=0$, the interior of the bubble with $R_0 = 60$ nm is perfect vacuum. An otherwise uniform plasma composed of cold ions and hot electrons characterised by $T_e = 1$ MeV, $Z = 1$, and $n_{i0} = n_{e0} = 3 \times 10^{21}$ cm$^{-3}$ are initiated. (**a**) Snapshots of the bubble collapse with the proton density distributions color-coded. The bubble is found to pulsate repeating implosion and explosion. (**b**) Temporal evolution of the proton energy spectrum for the same case as panel (a). At each collapse, proton flash is observed.

Figure 5 shows such a 2D result for a single bubble implosion. The 1$^{st}$, 2$^{nd}$, and 3$^{rd}$ row correspond to the ion density $n_i$, the absolute value of the electric field $E_f$ (magnified views), and the electron density $n_e$, respectively. The four columns correspond to different observation times, correlated with the laser intensity. In this simulation, a square-shaped target is normally irradiated from four directions by flat laser with the wavelength $\lambda_L = 1$ μm. The applied laser intensity on each side is given by $I_L(t) = I_0 \left\{ \frac{1}{2}(\cos(t/\tau_L) - 1) \right\}^2$ with the constants $I_0 = 5 \times 10^{18}$ W cm$^{-2}$ and $\tau_L = 20$ fsec. At $t = 0$, the plasma composed of cold electrons and protons with their initial density $n_{i0} = n_{e0} = 3 \times 10^{22}$ cm$^{-3}$ is initiated, while the bubble is set perfect vacuum in the same size as in Fig. 4. The generated hot electron temperature this time is self-consistently computed.

From Fig. 5, it can be seen that the electrons quickly fill the bubble rather uniformly owing to their high mobility and that the electric field around the imploding bubble consequently keeps its fully circular shape to drive the bubble-surface protons toward the center. In contrast to the bubble surface, the electric field with a speckled pattern in the solid is quickly smeared out with time. As a result, at a time immediately after the bubble collapse (see the upper right panel for $n_i$ at $t \simeq 51$ fsec), the formation of a nm-sized proton core is indeed observed at the center, where the full width at half maximum (FWHM) of the core profile turned out to be ~4 nm. This is already in the same order as the size of an unit cell. In other words, one needs even higher precision of the simulation to study the core dynamics for more details. Here it should be noted that the fine structures of shock propagation, formed in a Coulomb explosion of nanoscale clusters, were studied with high precision in refs[13,14]. Moreover, Peano et al.[15] studied a collisionless Coulomb explosion using a novel kinetic model to describe the electron dynamics. Their advanced numerical techniques are expected to be useful also for the study of bubble implosion.

Next we conducted 3D-MD simulation to quantitatively investigate the dynamics of the innermost protons, particularly when they converge at the centre. Recall that, in a spherically symmetric Coulomb system, a charged particle with radius $r$ is influenced only by particles contained in the spherical volume with radii smaller than $r$. Based on this fact, one can simulate the bubble implosion of a single atomic layer without taking the surrounding ions' influence into account. In the simulation we used $N_a = 10^3$ pseudoprotons (instead of $N_a \sim 10^8$ in a real system), which were initially arranged on a spherical surface at $R_0 = 1000\,nm$, while the electrons were treated as a uniform background corresponding to $\bar{n}_{e0} = 5 \times 10^{22}\,cm^{-3}$. In a real system, the electric field produced by the innermost protons should plateau due to the overwhelmingly large number of protons, $N_a \sim 10^8$, because the integrated effect of nonuniformity from all the surface protons is smeared as $N_a$ increases in proportion to $1/\sqrt{N_a}$. In other words, the nonuniformity level of the electric field in the MD simulation with $10^3$ particles is ~300 times larger than that of a real system. Thus, arranging such a small number of protons as uniformly as possible on a surface to mimic a real system is not trivial. To find the best initial configuration of the protons, we employed a self-organising method[35], which can achieve a lowest Coulomb potential energy of the surface proton system. The resultant configuration of charged particles provides the most uniform and most smooth self-field.

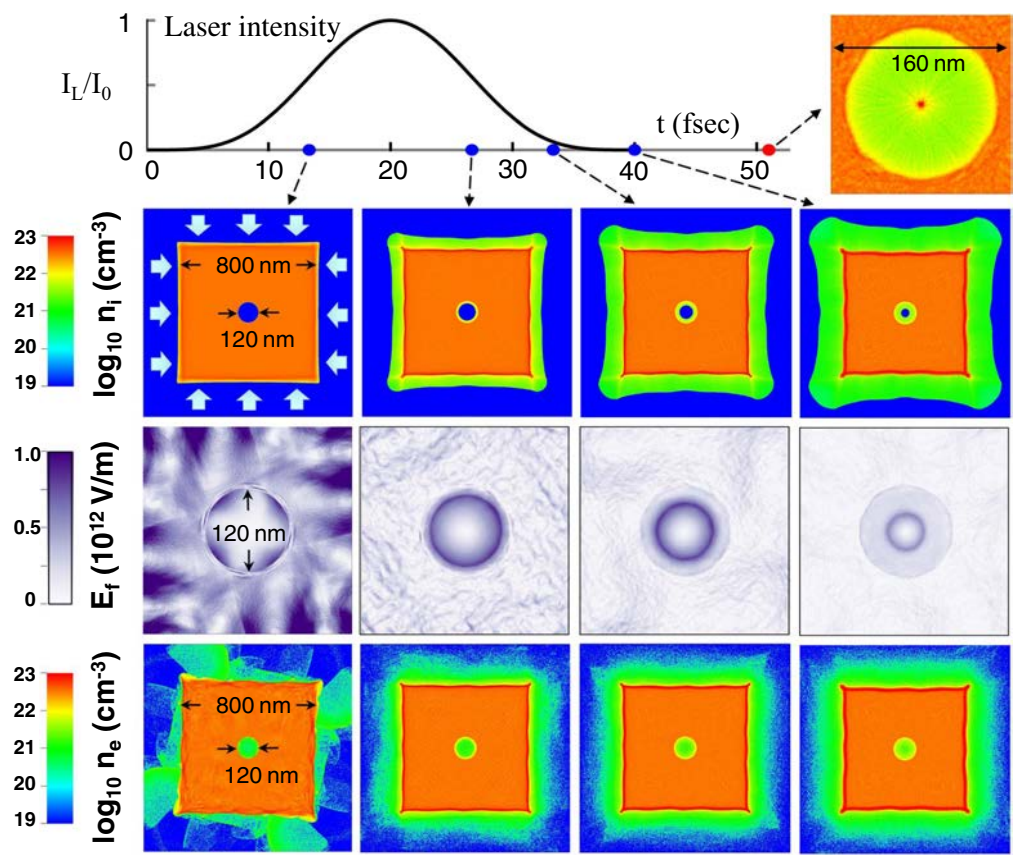

**Figure 5.** Two-dimensional ($x, y$) PIC simulation result for a single bubble implosion with realistic laser-matter interaction taken into account. All the physical quantities are assumed to be uniform along $z$-axis. At $t = 0$, the square-shaped target, composed of cold electrons and ions with their initial density $n_{i0} = n_{e0} = 3 \times 10^{22}\,cm^{-3}$, assigned with 60 pseudoparticles for each of ions and electrons in an unit cell, are initiated, while the interior of the bubble with $R_0 = 60\,nm$ is perfect vacuum. The target is then normally irradiated from four directions by flat laser with a wavelength $\lambda_L = 1\,\mu m$ and a peak intensity $I_0 = 5 \times 10^{18}\,W\,cm^{-2}$. Linearly polarized electric fields of the laser are assumed to be on the plane parallel to the square. Length of the calculation box, the target, and a single cell are 1200 nm, 800 nm, and 1 nm, respectively.

Figure 6(a) shows snapshots of the imploding particles obtained by the 3D-MD simulation, in which just the hemispherical domains are projected in four boxes with different scales ranging over three orders of magnitude. The original total mass and charge of $N_a \sim 10^8$ protons in a real system are kept unchanged with the $N_a = 10^3$ pseudoparticles. As a result, a single atomic layer can shrink to almost the same radius as that predicted by the simple model, i.e., $r_{min} \simeq 0.8\,nm$. The radial compression rate is $R_0/r_{min} \gtrsim 1000$, which is consistent with the result observed in Fig. 3(c). For comparison, the achievable radial compression ratios in other spherical convergent systems are, 30±40 in inertial confinement fusion[36] and 100±150 in sonoluminescence[29]. Hence, even a primitive system with a larger self-field nonuniformity than a real system can be compressed more than 1000 times.

Figure 6(b,c) show the temporal evolutions of the trajectories and the kinetic energies (interpreted for a real proton mass) for randomly sampled pseudoparticles, where the time origin of the horizontal axis is reset at the maximum compression for simplicity. The overlap of the maximum compression with the sample curves confirms the symmetric implosion. Note that the maximum exploding energy (Fig. 6(c)) is limited to twice the maximum imploding energy, which can be explained by the energy conservation law. Upon the maximum compression, when the pseudoparticles halt near their stagnating points, the trajectories become random. It should be noted

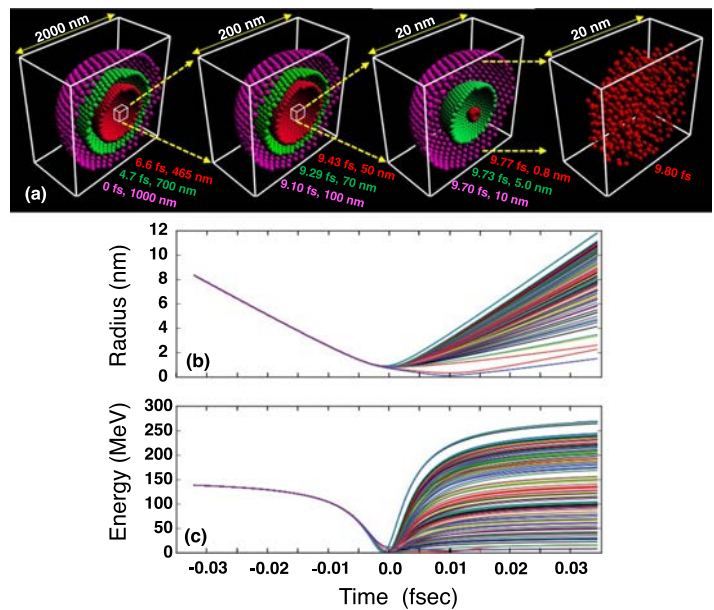

**Figure 6.** Three-dimensional molecular dynamic simulation of bubble implosion for a single atomic layer with $10^3$ pseudoparticles, that are initially arranged on a spherical surface at $R_0 = 1000$ nm, while the electrons are treated not as particles but as perfectly uniform background corresponding to $\bar{n}_{e0} = 5 \times 10^{22}$ cm$^{-3}$ assuming $\Lambda \ll 1$. The radial compression rate >1000 is coherent with the result observed in Fig. 3(c) at time E. (**a**) Snapshots of the imploding particles; time flows from left to right. The sets of measured time and averaged radius are given close to each snapshot. (**b**) Temporal evolution of trajectories for some pseudoparticles randomly sampled from ones in Fig. 6(a). (**c**) Temporal evolution of the kinetic energies (interpreted for a real proton mass). The time origin of the horizontal axis is reset to the maximum compression.

that, after the maximum compression is achieved, the physical picture given here does not make practical sense, because the innermost protons in the explosion phase interact with the protons just behind them, which is not considered in the current MD simulation.

In summary, we propose a novel concept, bubble implosion, to generate an ultrahigh field to accelerate protons to relativistic energies. A simple model and 1D, 2D, and 3D simulations comprehensively investigate the dynamics of the bubble implosion. This phenomenon is very likely to occur in reality. A stable implosion shrinks to a nanometre size and achieves an ultradense proton core, forming an unprecedentedly high electric field and producing proton flashes. The generation of an ultrahigh field is attributed to spherical convergence to the centre. Moreover, Coulomb-imploded bubbles are robust and behave as nano-pulsars repeating implosion and explosion to emit energetic protons. Although the present paper assumes pure hydrogen targets, a modified scenario should be applicable to other hydrides.

Current laser technology is suitable to experimentally identify bubble implosion by observing proton emissions at relativistic energies, which will be a major breakthrough to crack the 100-MeV barrier. For such experiments, a uniform and well-activated Coulomb field must be created inside the bubbles by laser irradiation of micron-sized bubbles embedded inside a solid target. We have demonstrated in terms of the 2D simulation that a symmetric bubble implosion can be achievable even under a realistic condition of laser-matter interaction. Consequently, the present concept should provide a new platform to elucidate fundamental phenomena in the fields of high-energy-density physics and astrophysics.

## References

1. Bulanov, S. V. & Khoroshkov, V. S. Feasibility of using laser ion accelerators in proton therapy. *Plasma Phys. Rep.* **28**, 453–456 (2002).
2. Malka, V. *et al.* Practicability of protontherapy using compact laser systems. *Med. Phys.* **31**, 1587–92 (2004).
3. Borghesi, M. *et al.* Fast ion generation by high-intensity laser irradiation of solid targets and applications. *Fusion Sci. Technol.* **49**, 412–439 (2006).
4. Roth, M. *et al.* Fast ignition by intense laser-accelerated proton beams. *Phys. Rev. Lett.* **86**, 436–439 (2001).
5. Krushelnick, K. *et al.* Ultrahigh-intensity laser-produced plasmas as a compact heavy ion injection source. *IEEE Trans. Plasma Sci.* **28**, 1184–1189 (2000).
6. Wilks, S. C. *et al.* Energetic proton generation in ultra-intense laser-solid interactions. *Phys. Plasmas* **8**, 542–549 (2001).
7. Snavely, R. A. *et al.* Intense high-energy proton beams from petawatt-laser irradiation of solids. *Phys. Rev. Lett.* **85**, 2945–2948 (2000).
8. Hatchett, S. P. *et al.* Electron, photon, and ion beams from the relativistic interaction of Petawatt laser pulses with solid targets. *Phys. Plasmas* **7**, 2076–2082 (2000).
9. Maksimchuk, A. *et al.* Forward ion acceleration in thin films driven by a high-intensity laser. *Phys. Rev. Lett.* **84**, 4108–4111 (2000).
10. Last, I. *et al.* Energetics and dynamics of Coulomb explosion of highly charged clusters. *J. Chem. Phys.* **107**, 6685–6692 (1997).
11. Ditmire, T. *et al.* High-energy ions produced in explosions of superheated atomic clusters. *Nature* **386**, 54–56 (1997).

12. Zweiback, J. *et al.* Nuclear fusion driven by Coulomb explosions of large deuterium clusters. *Phys. Rev. Lett.* **84**, 2634–2637 (2000).
13. Kaplan, A. E. & Dubetsky, B. Y. Shock shells in Coulomb explosions of Nanoclusters. *Phys. Rev. Lett.* **91**, 143401 (2003).
14. Peano, F. *et al.* Dynamics and control of shock shells in the Coulomb explosion of very large deuterium clusters. *Phys. Rev. Lett.* **94**, 033401 (2005).
15. Peano, F. *et al.* Kinetics of the Collisionless expansion of spherical nanoplasmas. *Phys. Rev. Lett.* **96**, 175002 (2006).
16. Murakami, M. & Mima, K. Efficient generation of quasimonoenergetic ions by Coulomb explosions of optimized nanostructured clusters. *Phys. Plasmas* **16**, 103108 (2009).
17. Esirkepov, T. *et al.* Highly efficient relativistic-ion generation in the laser-piston regime. *Phys. Rev. Lett.* **92**, 175003 (2004).
18. Henig, A. *et al.* Radiation-pressure acceleration of ion beams driven by circularly polarized laser pulses. *Phys. Rev. Lett.* **103**, 245003 (2009).
19. Kar, S. *et al.* Ion acceleration in multispecies targets driven by intense laser radiation pressure. *Phys. Rev. Lett.* **109**, 185006 (2012).
20. Yin, L. *et al.* Three-dimensional dynamics of breakout afterburner ion acceleration using high-contrast short-pulse laser and nanoscale targets. *Phys. Rev. Lett.* **107**, 045003 (2011).
21. Henig, A. *et al.* Enhanced laser-driven ion acceleration in the relativistic transparency regime. *Phys. Rev. Lett.* **103**, 045002 (2009).
22. Nakamura, T. *et al.* High-energy ions from near-critical density plasmas via magnetic vortex acceleration. *Phys. Rev. Lett.* **105**, 135002 (2010).
23. Bulanov, S. S. *et al.* Generation of GeV protons from 1 PW laser interaction with near critical density targets. *Phys. Plasmas* **17**, 043105 (2010).
24. Silva, L. O. *et al.* Proton shock acceleration in laser-plasma interactions. *Phys. Rev. Lett.* **92**, 015002 (2004).
25. Fiuza, F. *et al.* Laser-driven shock acceleration of monoenergetic ion beams. *Phys. Rev. Lett.* **109**, 215001 (2012).
26. Haberberger, D. *et al.* Collisionless shocks in laser-produced plasma generate monoenergetic high-energy proton beams. *Nat. Phys* **8**, 95–99 (2011).
27. Passoni, M. *et al.* Toward high-energy laser-driven ion beams: Nanostructured double-layer targets. *Phys. Rev. Accel. Beams* **19**, 061301 (2016).
28. Guderley, G. Starke kugelige und zylindrische Verdichtungsstöße in der Nähe des Kugelmittelpunktes bzw. der Zylinderachse. *Luftfahrtforschung* **19**, 302–312 (1942).
29. Barber, B. P. & Putterman, S. J. Light-scattering measurements of the repetitive supersonic implosion of a sonoluminescence bubble. *Phys. Rev. Lett.* **69**, 3839 (1992).
30. Rayleigh, Lord On the pressure developed in a liquid during the collapse of a spherical cavity. *Phil. Mag.* **34**, 94–98 (1917).
31. Nakamura, T. *et al.* Coulomb implosion mechanism of negative ion acceleration in laser plasmas. *Phys. Lett. A* **373**, 2584–2587 (2009).
32. Nakamura, T. *et al.* High energy negative ion generation by Coulomb implosion mechanism. *Phys. Plasmas* **16**, 113106 (2009).
33. Son, S. & Fisch, N. J. Pycnonuclear reaction and possible chain reactions in an ultra-dense DT plasma. *Phys. Lett. A* **337**, 397–407 (2005).
34. Arber, T. D. *et al.* Contemporary particle-in-cell approach to laser-plasma modeling. *Plasma Phys. Control. Fusion* **57**, 113001 (2015).
35. Murakami, M. *et al.* Optimization of irradiation configuration in laser fusion utilizing self-organizing electrodynamic system. *Phys. Plasmas* **17**, 082702 (2010).
36. Lindl, J. D. *Inertial Confinement Fusion*. (Springer, New York, 1998).

### Acknowledgements
This work was supported by the Japan Society for the Promotion of Science (JSPS). Simulations were performed using the EPOCH code (developed under UK EPSRC Grants No. EP/G054940/1, No. EP/G055165/1, and No. EP/G056803/1) using HPC resources provided by the TACC at the University of Texas and the Comet cluster at the SDSC at the University of California at San Diego. This work used the Extreme Science and Engineering Discovery Environment (XSEDE), which is supported by National Science Foundation grant number ACI-1548562. One of the authors (M.M.) thanks Prof. Y. Sentoku for a trial simulation.

### Author Contributions
M.M. conceived the physical idea, performed 1D simulation, and developed the simple model. A.A. and M.A.Z. performed the PIC and MD simulations, respectively. M.M. wrote the paper. All authors reviewed the whole work, and approved the manuscript.

### Additional Information
**Competing Interests:** The authors declare no competing interests.

**Publisher's note:** Springer Nature remains neutral with regard to jurisdictional claims in published maps and institutional affiliations.